\documentclass{article}

\usepackage{microtype}
\usepackage{graphicx}
\usepackage{booktabs}   
\usepackage{amsmath,amssymb}
\usepackage{url}
\usepackage{hyperref}

\usepackage[accepted]{mlsys2026}
\mlsystitlerunning{A Training-Memory Regression in MLA Sequence Parallelism}

\newcommand{\laga}{\textsc{Laga}}
\newcommand{\assertMLA}{\texttt{assert\allowbreak\ not\allowbreak\ (self.\allowbreak training\allowbreak\ and\allowbreak\ self.\allowbreak cache\_\allowbreak mla\_\allowbreak latents)}}

\begin{document}

\twocolumn[
\mlsystitle{A Training-Memory Regression in MLA Sequence Parallelism:\\
Why Megatron-Core Forbids Absorption, and \laga{} --- a Communication-Efficient Fix}

\begin{mlsysauthorlist}
\mlsysauthor{Changzheng Ma}{cm}
\end{mlsysauthorlist}
\mlsysaffiliation{cm}{China Mobile Jiutian Artificial Intelligence Technology (Beijing) Co., Ltd., Beijing, China}
\mlsyscorrespondingauthor{Changzheng Ma}{machangzheng@cmjt.chinamobile.com}
\mlsyskeywords{Multi-head Latent Attention, Sequence Parallelism, Distributed Training, MoE}

\vskip 0.3in

\begin{abstract}
Multi-head Latent Attention (MLA) ships two implementations in Megatron-Core:
an \emph{explicit} form used for training and an \emph{absorbed} form ---
which slashes collective communication by gathering only the compressed
latent --- that is fully implemented but \textbf{hard-asserted out of
training} (\assertMLA),
allowed only in inference decode. The library documents no reason. \textbf{We
show the restriction is well-founded and quantify why:} ported to training,
the absorbed form is a \emph{memory trap} --- its intermediates live in
$n_h\!\cdot\!d_{kv}$ dimensions per token, larger than the per-head K/V they
replace --- inflating activation memory by \textbf{20--34\%, up to
9.2\,GB at DeepSeek-V3 scale} ($n_h{=}128$, $\textrm{seq}{=}16384$,
$\textrm{SP}{=}8$, eager kernel; the gap widens to \textbf{19.2\,GB} under a
fused kernel, \S\ref{sec:fused}), enough to change device-fit. This measurement, validated
on two axes (linear in seq and $n_h$) and cross-verified on NVIDIA A100,
explains the otherwise-undocumented restriction and leaves practitioners with
no low-communication MLA training path. \textbf{We then provide one.}
\laga{} (Latent All-Gather Attention) keeps the absorbed form's latent-gather
communication but rejects the absorb reformulation, instead reconstructing
per-head K/V locally from the gathered latent. On $8\times$Ascend 910B at
real DeepSeek-V3 dimensions, \laga{} cuts collective communication
\textbf{1.98$\times$}, matches explicit memory within \textbf{$\leq$0.5\%}, is
\textbf{bit-identical} to explicit at SP=1 and equivalent to
${<}1{\times}10^{-3}$ at SP=2--8, and under a fused attention kernel improves
attention-block throughput \textbf{1.04--1.06$\times$ single-node and
1.07--1.24$\times$ cross-node} --- leading at all sequence lengths in the
cross-node regime MLA is deployed for.
\end{abstract}
]

\printAffiliationsAndNotice{}

\section{Introduction}
\label{sec:intro}

\noindent\textbf{A puzzle in production code.} Open Megatron-Core's MLA
implementation~\cite{megatroncore} and the forward begins with an assertion
that bans its own low-communication path from training:
\assertMLA. The absorb
reformulation --- folding \texttt{kv\_b\_proj} into the query so attention
runs against the compressed latent and only the small latent crosses the
collective --- is fully implemented in the class, but gated to
\texttt{is\_decode\_only()} inference and hard-asserted out of training. The
library ships no low-communication MLA \emph{training} path: training must
use the explicit form, which all-to-alls per-head K/V and at DeepSeek-V3's
$n_h{=}128$ dominates communication. Why does Megatron-Core refuse to absorb
during training? The library does not say.

\noindent\textbf{We answer the puzzle, and quantify the answer.} Ported to
training, the absorbed form is a \emph{memory trap}. Its intermediates ---
$q_{\textrm{absorbed}} = W_{kv\_b}^{\top} q_{\textrm{nope}}$ and the
post-attention latent accumulator --- live in $n_h\!\cdot\!d_{kv}$ dimensions
per token, \emph{larger} than the per-head K/V they replace whenever
$d_{kv} > d_{\textrm{head}}$. At DeepSeek-V3 scale ($n_h{=}128$,
$d_{\textrm{model}}{=}7168$, $\textrm{seq}{=}16384$, $\textrm{SP}{=}8$) we
measure the absorbed form's peak activation memory exceeding the explicit
form by \textbf{20--34\% --- 9.2\,GB at the production-relevant
configuration} --- enough to change whether the model fits on device. The
absorb trick is an \emph{inference} KV-cache optimization (store the latent,
not per-head KV, a $\approx 32\times$ cache win); in training there is no
cache, every forward recomputes, and the absorbed intermediates cost memory
rather than save it. Megatron-Core's \texttt{assert} is, in effect, an
undocumented guard against this regression.

\noindent\textbf{A training path, as a consequence.} The measurement above
leaves production with a gap: the only low-communication MLA pattern
(absorbed) is training-incompatible, so training must use the
communication-heavy explicit path. The absorb reformulation is unnecessary
for closing this gap --- what we actually want is to stop \emph{moving
per-head K/V across the collective}, and MLA already provides a compressed
latent to move instead. \laga{} keeps the explicit attention math (no absorb,
no latent-dimension intermediates) and changes only \emph{what crosses the
wire}: a cheap latent all-gather replaces the expensive K/V all-to-all, and
per-head K/V are reconstructed \emph{locally} on each rank from the gathered
latent, at head-shard granularity. \laga{} inherits the absorbed form's
communication and the explicit form's memory.

\begin{figure*}[t]
\centering
\includegraphics[width=\textwidth]{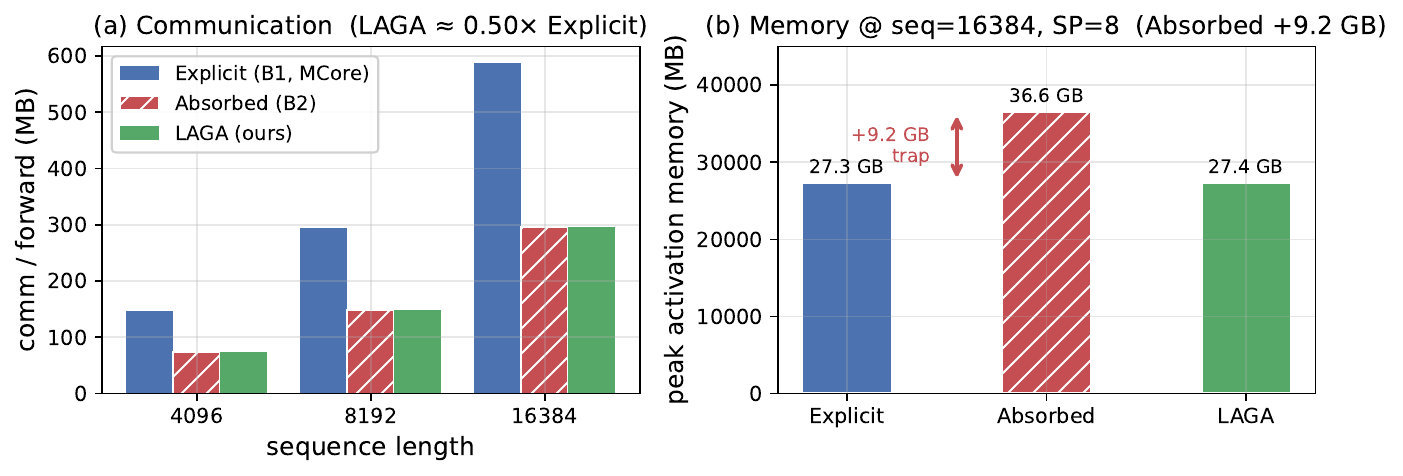}
\caption{\textbf{Teaser.} Left: per-forward communication at DeepSeek-V3 scale
(explicit $\to$ \laga{}/absorbed $\approx 50\%$, 1.98$\times$). Right: peak
activation memory --- \textbf{absorbed $+$9.2\,GB at seq=16384/SP=8} (the trap
we measure and explain; $+$19.2\,GB under fused attention without the eager
score-matrix confound), \textbf{\laga{} = explicit} (the consequent fix).
The headline finding is the right panel's memory trap and its cause;
\laga{} is the training path that removes it while keeping the left panel's
1.98$\times$.}
\label{fig:teaser}
\end{figure*}

\noindent\textbf{Background: why MLA $+$ SP is hard.} Long-context training
depends on sequence parallelism; the dominant head-dimension family
(DeepSpeed Ulysses~\cite{jacobs2023ulysses} and descendants) repartitions
seq$\leftrightarrow$head so each rank attends full-sequence for a subset of
heads, assuming each rank owns a per-head KV shard to scatter. Multi-head
Latent Attention~\cite{deepseekv2,deepseekv3} compresses KV into a single
shared low-rank latent ($d_{kv}\approx 512$, vs $n_h\!\cdot\!d_{\textrm{head}}
\approx 16384$ for equivalent GQA), so there is no per-head KV to scatter
--- in the absorbed form the effective Ulysses degree collapses to one. This
is the structural reason production lands on the communication-heavy explicit
path.

\noindent\textbf{Contributions.}
\begin{itemize}\itemsep0pt
\item \textbf{C1 --- The absorbed-memory-trap: measurement, mechanism, and
explanation (\S\ref{sec:trap}/\ref{sec:mem}).} By porting the inference-side
absorb trick to training, we quantify a 20--34\% activation-memory inflation
--- reaching \textbf{9.2\,GB at DeepSeek-V3 scale} ($n_h{=}128$,
seq$=$16384, SP$=$8, eager kernel; \textbf{19.2\,GB} under a fused kernel,
\S\ref{sec:fused}); trace its \emph{cause} to
$n_h\!\cdot\!d_{kv}$-dim intermediates exceeding the per-head K/V they
replace; and show the inflation is \emph{linear on two independent axes}
(sequence length and $n_h$), cross-verified on NVIDIA A100. This explains
Megatron-Core's otherwise-undocumented
\assertMLA.
\item \textbf{C2 --- \laga{}, the consequent training fix
(\S\ref{sec:laga}).} A low-communication MLA training path --- latent
all-gather $+$ local head-shard up-projection, no absorb --- that closes the
gap C1 exposes. Not present in Megatron-Core.
\item \textbf{C3 --- Three-way characterization $+$ crossover
(\S\ref{sec:eval}).} Closed-form communication/memory analysis validated at
V3 dims on $8\times$Ascend 910B; a throughput crossover showing \laga{}'s
advantage \emph{strengthens at production scale} (wins single-node at
$n_h{=}128$) and whose sequence-driven split sharpens cross-node, where long
context is runnable at \textbf{$+$24\%} under fused attention, with \laga{}
leading at all sequence lengths cross-node.
\item \textbf{C4 --- Correctness (\S\ref{sec:correctness}).} Bit-identical
to explicit at SP=1 (same matmul, reordered views); equivalent to
${<}7{\times}10^{-4}$ at SP=2--8; verified by a multi-step convergence run.
\end{itemize}

\section{Background}
\label{sec:bg}

\subsection{Multi-head Latent Attention (MLA)}
MLA~\cite{deepseekv2} factorizes KV to reduce memory. Given hidden
$h\in\mathbb{R}^{B\times S\times d}$:
\begin{align*}
q       &= W_q\,h,\;\; c = W_{kv\_a}\,h\\
k_{pe}  &= \textrm{split}_{1}(c),\;\;
          \textrm{latent} = \textrm{split}_{2}(c)\in\mathbb{R}^{B\times S\times d_{kv}}\\
kv      &= W_{kv\_b}\,\textrm{LN}(\textrm{latent}),\;\;
          (k_{\textrm{nope}},\,v) = \textrm{split}(kv)\\
\textrm{scores} &= (q_{\textrm{nope}}\!\parallel\! q_{\textrm{rope}})
                   (q_{\textrm{nope}}\!\parallel\! k_{pe})^{\top}\!/\sqrt{d}
\end{align*}
(The latent is shared across all $n_h$ heads; $k_{pe}$ is broadcast.)
For DeepSeek-V3 dims $n_h{=}128$, $d_{\textrm{nope}}{=}d_v{=}128$,
$d_{\textrm{rope}}{=}64$, $d_{kv}{=}512$, $d_{\textrm{model}}{=}7168$, the
latent is $n_h\!\cdot\!d_{\textrm{head}}/d_{kv}\approx 32\times$ smaller
than equivalent per-head KV.
Two properties matter for SP. \textbf{(P1) Shared latent:} the K/V of
\emph{all} heads descend from a single $d_{kv}$-rank tensor.
\textbf{(P2) Decoupled RoPE:} a small $d_{\textrm{rope}}$-dim shared band
$k_{pe}$ carries positional information separately.

\subsection{Sequence parallelism}
SP partitions the sequence so each rank attends to a sub-sequence. The
head-dimension family --- Megatron-SP~\cite{korthikanti2023megatronsp},
Ulysses~\cite{jacobs2023ulysses}, Ring
Attention~\cite{liu2024ringattention}, USP~\cite{usp2024},
LoongTrain~\cite{loongtrain2024} --- repartitions seq$\leftrightarrow$head so
each rank computes full-sequence attention for $n_h/P$ heads. \textbf{All
assume each rank owns a per-head KV shard to scatter} --- true for MHA/GQA,
false under MLA property P1.

\subsection{Megatron-Core's MLA path (and why absorption is inference-only)}
\label{sec:mcore}
Megatron-Core (the de-facto reference production stack, building on
Megatron-LM~\cite{narayanan2021megatronlm}) ships a single MLA training
class, \texttt{MLASelfAttention}. Its training forward is the
\textbf{explicit} form: \texttt{linear\_kv\_up\_proj} materializes
per-head K/V \emph{before} \texttt{core\_attention}, which carries per-head
K/V over the CP collective. MCore thus trains MLA long-context under the
communication-heavy explicit pattern.

MCore \emph{also} contains the absorb reformulation --- but it is
\textbf{gated to inference-decode and hard-asserted against training}. The
forward opens with
\begin{verbatim}
assert not (self.training
            and self.cache_mla_latents)
\end{verbatim}
and the absorption itself,
\texttt{q\_content = torch.einsum(\ldots)} folding \texttt{kv\_b\_proj}
into the query, runs only under the guard
\texttt{use\_absorption = cache\_mla\_latents \&\&
inference\_context.is\_decode\_only()}. MCore's
designers deliberately restrict the absorb trick to inference: the class's
\texttt{prepare\_for\_absorption} even concedes that it is \emph{``not doing
true absorption. We will add this support at a later time.''} \textbf{One
contribution of this paper (\S\ref{sec:trap}/\ref{sec:mem}) is to explain
why that restriction is well-founded} --- the absorbed form, ported to
training, regresses activation memory by 20--34\% (up to 9.2\,GB at V3 scale)
--- and to offer \laga{}, a latent-over-CP path viable for training, which
MCore does not ship.

\section{\laga{}: Latent All-Gather Attention}
\label{sec:laga}

We present \laga{}, a training-time sequence parallelism for MLA. We first
pinpoint why standard Ulysses-style SP fails on MLA (\S\ref{sec:whyfail}),
then analyze two natural-but-flawed approaches (\S\ref{sec:trap}), and
finally present \laga{} (\S\ref{sec:lagaalgo}). Throughout, \textbf{Explicit
(B1)} denotes the explicit Ulysses-on-MLA baseline, \textbf{Absorbed (B2)}
the inference-side absorb trick ported to training, and \textbf{\laga{}} our
method.

\subsection{Why MLA breaks head-dimension SP}
\label{sec:whyfail}
Ulysses-style SP~\cite{jacobs2023ulysses} partitions the \emph{sequence}
dimension and repartitions onto the \emph{head} dimension via an all-to-all,
so each rank computes full-sequence attention for a subset of heads. This
requires every rank to own at least one KV head to scatter. MLA defeats this
assumption: its KV is a \textbf{single shared low-rank latent}
$\textrm{latent}_{kv}\in\mathbb{R}^{d_{kv}}$ (plus a decoupled RoPE band
$k_{pe}$), broadcast across all heads via $W_{kv\_b}$. There is no per-head
KV to scatter along the head axis. In the \emph{absorbed} formulation the K
side is a shared $d_{kv}$-rank tensor, so the maximum Ulysses SP degree
collapses to 1.

The only way to apply head-dimension SP to MLA is to \textbf{materialize
per-head K/V first} (the \emph{explicit} form): compute
$kv = W_{kv\_b}\,\textrm{LN}(\textrm{latent}) \to$ per-head $k_{\textrm{nope}}, v$,
then all-to-all them along heads. This works, but the K/V all-to-all moves
$n_h\!\cdot\!(d_{\textrm{nope}}+d_v)$ elements per token --- expensive when
$n_h$ is large (DeepSeek-V3: 128 heads). \textbf{\laga{}'s goal: keep
head-dimension SP, but replace the expensive per-head K/V all-to-all with a
cheap latent all-gather --- without the memory penalty of the naive absorbed
approach.}

\subsection{Two natural baselines (and why they fall short)}
\label{sec:trap}
\noindent\textbf{Explicit Ulysses-on-MLA (B1) --- the production training
path.} Materialize per-head K/V, then all-to-all Q, K, V along heads.
Correct, but communication scales with $n_h\!\cdot\!d_{\textrm{head}}$. This
is exactly Megatron-Core's MLA training path.

\noindent\textbf{Absorbed SP (B2) --- the inference-side absorb trick, ported
to training.} Absorb $W_{kv\_b}$'s K-half into the query ($q_{\textrm{absorbed}}$,
an $n_h\!\cdot\!d_{kv}$-rank tensor), gather only the latent across CP, run
attention in latent space, apply V's up-projection post-CP. Communication
drops to that of \laga{}. \textbf{But this is a training-memory trap:} the
intermediates $q_{\textrm{absorbed}}$ and the post-attention latent
accumulator live in $n_h\!\cdot\!d_{kv}$ dimensions per token, \emph{larger}
than the per-head K/V they replace whenever $d_{kv} > d_{\textrm{head}}$.
Measured at DeepSeek-V3 scale ($n_h{=}128$, $d_{kv}{=}512$,
$d_{\textrm{head}}{=}128$), B2's peak activation memory exceeds B1 by
\textbf{20--34\%, reaching $+$9.2\,GB at seq=16384/SP=8} (Table~\ref{tab:mem})
--- enough to change device-fit. This both explains MCore's
training-restriction on absorption and motivates \laga{}.

\subsection{\laga{}: latent all-gather + local up-projection}
\label{sec:lagaalgo}
\laga{} keeps B2's communication pattern (all-gather the latent) but rejects
the absorb reformulation, instead \textbf{reconstructing per-head K/V
locally} from the gathered latent --- landing them in the head dimension
(like B1), not the latent dimension (like B2). Algorithm~\ref{alg:laga} gives
the forward pass for SP degree $P$.

\begin{algorithm*}[t]
\caption{\laga{} attention forward (rank $r$, SP degree $P$).}
\label{alg:laga}
\centering
\fbox{\begin{minipage}{0.97\textwidth}\small
\emph{Input:} local hidden $h\in\mathbb{R}^{B\times (S/P)\times d}$. Per-rank
head shard $H_r=\{r\,n_h/P,\ldots,(r{+}1)n_h/P{-}1\}$.
\begin{enumerate}\itemsep0pt
\item $q_{\textrm{nope}}, q_{\textrm{rope}} \leftarrow W_q(h)$;\;
       $\textrm{latent}, k_{pe} \leftarrow W_{kv\_a}(h)$;\;
       $\textrm{latent}\leftarrow \textrm{LayerNorm}(\textrm{latent})$. \hfill(local)
\item $q_{\textrm{nope}}, q_{\textrm{rope}} \leftarrow
       \textrm{AllToAll}_{\textrm{head}}(q_{\textrm{nope}}, q_{\textrm{rope}})$
       --- head-scatter, gather seq. \hfill(cheap: $n_h(d_{\textrm{nope}}+d_{\textrm{rope}})$ elem/tok)
\item $\textrm{latent}, k_{pe} \leftarrow \textrm{AllGather}_{\textrm{seq}}(\textrm{latent}, k_{pe})$
       --- gather along seq. \hfill\textbf{key: $d_{kv}+d_{\textrm{rope}}$ elem/tok, vs $n_h(d_{\textrm{nope}}+d_v)$ for B1's K/V a2a}
\item $k_{\textrm{nope}} \leftarrow W_{kv\_b}^{k}[H_r]\!\cdot\!\textrm{latent}$;\quad
       $v \leftarrow W_{kv\_b}^{v}[H_r]\!\cdot\!\textrm{latent}$
       --- \textbf{local up-projection of this rank's head shard only}.\hfill(head-dim output, no latent-dim inflation)
\item $\textrm{scores}\leftarrow(q_{\textrm{nope}}\!\parallel\!q_{\textrm{rope}})
       (k_{\textrm{nope}}\!\parallel\!k_{pe})^{\top}/\sqrt{d}$;\quad
       $\textrm{out}\leftarrow\textrm{softmax}(\textrm{scores})\,v$
\item $\textrm{out}\leftarrow\textrm{AllToAll}_{\textrm{seq}}(\textrm{out})$
       --- scatter seq, gather heads $\to\mathbb{R}^{B\times(S/P)\times n_h\times d_v}$
\item \textbf{return} $W_o(\textrm{reshape}(\textrm{out}))$
\end{enumerate}
\end{minipage}}
\end{algorithm*}

\noindent\textbf{Why this is both comm-cheap and memory-neutral.} Step~3
moves $d_{kv}+d_{\textrm{rope}}$ elements per token instead of B1's
$n_h(d_{\textrm{nope}}+d_v)$ --- yielding \textbf{1.98$\times$ less total
forward communication at V3 scale} (Table~\ref{tab:comm}). Step~4 produces
K/V in the head dimension at head-shard granularity
$(n_h/P)(d_{\textrm{nope}}+d_v)$ --- identical to B1's post-all-to-all
footprint, and \emph{smaller than B2's $n_h d_{kv}$
$q_{\textrm{absorbed}}$/accumulator}. Measured: \laga{} peak memory matches
B1 within $\leq$0.5\% ($<$27\,MB) while B2 balloons by up to 9.2\,GB
(Table~\ref{tab:mem}).

\noindent\textbf{Correctness.} \laga{} is mathematically identical to B1:
step~4's $W_{kv\_b}^{k}[H_r]\!\cdot\!\textrm{latent}$ is exactly the $H_r$
slice of B1's $W_{kv\_b}(\textrm{latent})$ (same parameters, same matmul,
different view order). Under SP=1 the two are \textbf{bit-identical} (output
$\max|\Delta|=0.000$; \S\ref{sec:correctness}). The $W_{kv\_b}$ gradient
shards along the head axis under \laga{}, so backward requires a head-axis
gradient all-reduce --- a one-line implementation detail
(\S\ref{sec:backward}).

\subsection{Composition with expert parallelism}
\laga{}'s collectives live inside the attention block (head/seq axes);
expert parallelism's dispatch all-to-all lives inside the MoE block (token
axis). The two are temporally separated and compose without conflict
(\S\ref{sec:ep}).

\begin{figure*}[t]
\centering
\includegraphics[width=\textwidth]{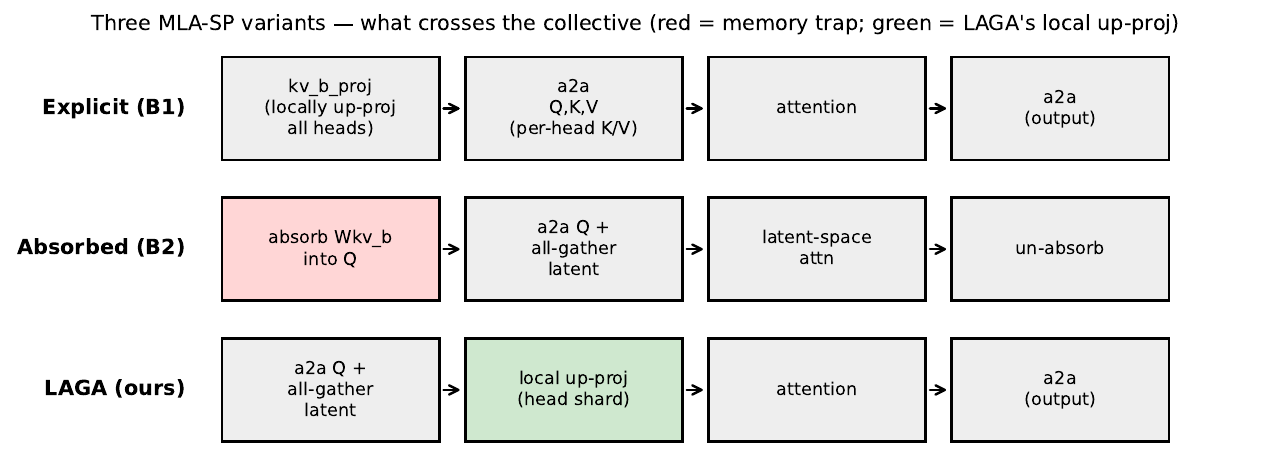}
\caption{\textbf{Dataflow.} Explicit (B1): $W_{kv\_b}\!\to\!a2a(Q,K,V)\!\to\!\textrm{attn}\!\to\!a2a$.
Absorbed (B2): $a2a(Q)+ag(\textrm{latent})\!\to\!\textrm{absorb}\!\to\!\textrm{latent-attn}\!\to\!\textrm{un-absorb}\!\to\!a2a$.
\textbf{\laga{}}: $a2a(Q)+ag(\textrm{latent})\!\to\!\textrm{local up-proj}(K,V)\!\to\!\textrm{attn}\!\to\!a2a$.
\laga{} shares B2's left and right spine but replaces latent-space attention
with head-dim attention on locally reconstructed K/V.}
\label{fig:dataflow}
\end{figure*}

\section{Evaluation}
\label{sec:eval}

We evaluate \laga{} against \textbf{Explicit (B1)} and \textbf{Absorbed (B2)}
on communication volume (\S\ref{sec:comm}), activation memory
(\S\ref{sec:mem}), and throughput (\S\ref{sec:tp}), then verify numerical
equivalence to B1 (\S\ref{sec:numeq}).

\subsection{Setup}
\label{sec:setup}
\noindent\textbf{Hardware.} One node of $8\times$Ascend 910B (HCCL) for
single-node experiments, and \textbf{two nodes $\times$ 8 cards (16 ranks)}
for the multi-node SP group (spanning both nodes over RoCE $\approx
11$\,GB/s). All measurements are bf16. \textbf{Structural results and the
throughput crossover shape are cross-validated on NVIDIA A100
(\S\ref{sec:xhw}).}

\noindent\textbf{Model (real DeepSeek-V3 dims).} A standalone MLA transformer
with public DeepSeek-V3 dimensions: $d_{\textrm{model}}{=}7168$, $n_h{=}128$,
$d_{\textrm{nope}}{=}d_v{=}128$, $d_{\textrm{rope}}{=}64$, $d_{kv}{=}512$,
MLP intermediate $=8192$. $S\in\{4096,8192,16384\}$. Comm/memory use a single
attention layer; throughput uses a 4-layer stack at a smaller $n_h$ config
(see \S\ref{sec:tp}). The implementation is eager PyTorch (no fused attention
kernels); \S\ref{sec:limits} discusses this scope.

\noindent\textbf{Baselines.} B1 and B2 share the \emph{same} weights, input,
and SP configuration as \laga{} --- only the attention form differs
(controlled comparison). B1 reproduces the \emph{mathematics} of
Megatron-Core's \texttt{MLASelfAttention} explicit path (the production
training form); our comm/memory claims are analytical or structural and thus
implementation-independent, while throughput numbers are from our prototype
(not measured against MCore directly, since MCore's absorbed path is
hard-asserted against training and the CUDA stack differs on NPU --- see
\S\ref{sec:limits}). Comm bytes counted as per-rank send volume.

\subsection{Communication volume (V3 scale, $n_h{=}128$)}
\label{sec:comm}
\begin{figure}[t]
\centering
\includegraphics[width=\columnwidth]{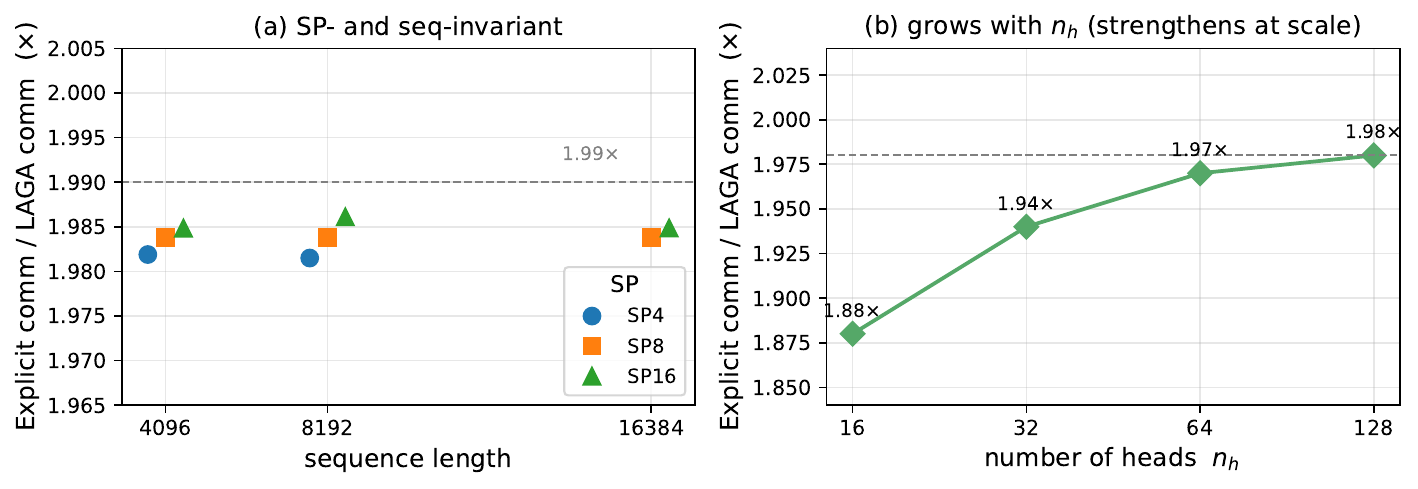}
\caption{\textbf{Communication reduction is structural and scale-dependent.}
(a) Explicit/\laga{} ratio holds $\approx1.98\times$ across SP and seq.
(b) It climbs $1.88\!\to\!1.98\times$ as $n_h$ grows $16\!\to\!128$ ---
strengthens at production scale.}
\label{fig:comm}
\end{figure}

Table~\ref{tab:comm} reports per-rank, per-forward sent bytes.
\laga{} and B2 move \textbf{identical} data; both cut communication to
\textbf{50.5\% of B1} --- a stable \textbf{1.98$\times$ reduction} across
SP=4--8 and $S$=4K--16K. The ratio is structural: B1's K-side all-to-all
moves $n_h(d_{\textrm{nope}}+d_v)$ elements/token, whereas \laga{} moves the
low-rank $d_{kv}+d_{\textrm{rope}}$. The reduction \emph{strengthens with
$n_h$}: sweeping $n_h\in\{16,32,64,128\}$ at SP=8/$S$=8192, the ratio climbs
\textbf{1.88$\times\!\to\!1.94\!\to\!1.97\!\to\!1.98$} (Fig.~\ref{fig:comm}b)
--- exactly as predicted, since only B1's K-side term grows with $n_h$.

\begin{table*}[t]
\caption{Per-rank per-forward communication (MB), DeepSeek-V3 dims ($n_h{=}128$, $d{=}7168$).
\laga{}/B2 $\approx 0.50\times$ B1 (1.98$\times$).}
\label{tab:comm}
\centering\footnotesize
\begin{tabular}{c|ccc|ccc|ccc}
\toprule
& \multicolumn{3}{c|}{$S{=}4096$} & \multicolumn{3}{c|}{$S{=}8192$} & \multicolumn{3}{c}{$S{=}16384$}\\
SP & B1 & B2 & \laga{} & B1 & B2 & \laga{} & B1 & B2 & \laga{}\\
\midrule
4  & 251.7 & 127.0 & \textbf{127.0} & 503.3 & 254.0 & \textbf{254.0} & --- & --- & ---\\
8  & 146.8 & 74.0  & \textbf{74.0}  & 293.6 & 148.0 & \textbf{148.0} & 587.2 & 296.0 & \textbf{296.0}\\
16 & 78.6  & 39.6  & \textbf{39.6}  & 157.3 & 79.2  & \textbf{79.2}  & 314.6 & 158.5 & \textbf{158.5}\\
\bottomrule
\end{tabular}
\end{table*}

\subsection{Activation memory (V3 scale) --- the absorbed memory trap}
\label{sec:mem}
Table~\ref{tab:mem} reports peak \texttt{max\_memory\_allocated}.
\textbf{\laga{} matches B1 within $\leq$0.5\%}, while \textbf{B2 inflates memory
by 20--34\% --- growing sharply with sequence length}, reaching
\textbf{$+$9.2\,GB at $S{=}16384$/SP=8}. This is the central empirical
finding: the absorbed path's $q_{\textrm{absorbed}}$/accumulator
intermediates (in $n_h d_{kv}=128\!\cdot\!512$ per token) dominate as
sequences grow. The inflation also scales \textbf{linearly with $n_h$} ---
337/657/1269/2512\,MB as $n_h$ doubles $16\!\to\!128$ (at $S{=}8192$/SP=8;
the sweep uses $d_{\textrm{model}}{=}2048$, so the $n_h{=}128$ point's
2512\,MB differs from the 2362\,MB at $d_{\textrm{model}}{=}7168$ in
Table~\ref{tab:mem} --- the intermediate scales with $d_{\textrm{model}}$)
--- confirming the $n_h d_{kv}$-intermediate mechanism on a second axis.

\begin{figure}[t]
\centering
\includegraphics[width=\columnwidth]{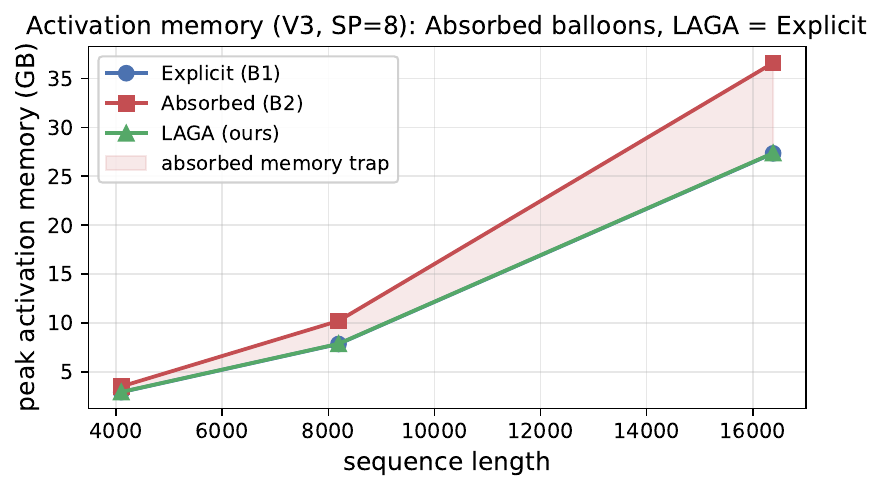}
\caption{Activation memory (V3, SP=8): Absorbed balloons with sequence
length; \laga{} tracks Explicit. Shaded gap = the absorbed memory trap.}
\label{fig:mem}
\end{figure}

\begin{table*}[t]
\caption{Peak activation memory (MB), DeepSeek-V3 dims. \laga{} = B1 ($<$0.4\%);
B2 balloons up to $+$9.2\,GB.}
\label{tab:mem}
\centering\footnotesize
\begin{tabular}{c|ccc|ccc|ccc}
\toprule
& \multicolumn{3}{c|}{$S{=}4096$} & \multicolumn{3}{c|}{$S{=}8192$} & \multicolumn{3}{c}{$S{=}16384$}\\
SP & B1 & B2 & \laga{} & B1 & B2 & \laga{} & B1 & B2 & \laga{}\\
\midrule
4 & 4620 & 5900 & \textbf{4633} & 14463 & 19409 & \textbf{14485} & --- & --- & ---\\
8 & 2917 & 3493 & \textbf{2929} & 7842  & 10207 & \textbf{7860}  & 27348 & 36596 & \textbf{27375}\\
\bottomrule
\end{tabular}
\\[2pt]
\raggedright\footnotesize B2 inflation: +1.3\,/\,4.9\,GB at SP=4; +0.6\,/\,2.4\,/\,\textbf{9.2}\,GB at SP=8 for $S{=}4$K/8K/16K.
\end{table*}

\subsection{Throughput}
\label{sec:tp}
We report tokens/s for forward+backward. The picture is a clean
\textbf{crossover}, sharpened by multi-node topology.

\noindent\textbf{Single-node (SP=8, smaller $n_h{=}32$, Table~\ref{tab:tp32}).}
In the communication-bound regime (long $S$) \laga{} reaches \textbf{parity}
(1.01$\times$) at $S{=}16384$; in the compute-bound regime (short $S$) \laga{}
is $\approx$21\% slower. Note \laga{}'s \emph{total FLOPs equal} B1's
($W_{kv\_b}$ is merely redistributed) --- the regression is \emph{kernel
efficiency}, not FLOPs: \laga{}'s per-head-shard einsums are smaller GEMMs.
B2 is consistently slowest.

\begin{table*}[t]
\caption{Single-node throughput (tokens/s, SP=8, $n_h{=}32$).
Each cell reports \textbf{B1 / B2 / \laga{}}; bold marks the best of the three.
Last column: \laga{}/B1 ratio. ``---'' = not run at that configuration.}
\label{tab:tp32}
\centering\footnotesize
\setlength{\tabcolsep}{5pt}
\begin{tabular}{c|ccc|c}
\toprule
SP & $S{=}4096$ & $S{=}8192$ & $S{=}16384$ & \laga{}/B1\\
 & (B1 / B2 / \laga{}) & (B1 / B2 / \laga{}) & (B1 / B2 / \laga{}) & \\
\midrule
2 & 24837 / 18793 / \textbf{27790} & 17708 / 11564 / \textbf{19341} & --- & 1.12 / 1.09\\
4 & 47663 / 35521 / 46353 & 36764 / 23247 / \textbf{38217} & --- & 0.97 / 1.04\\
8 & 49388 / 35411 / 39248 & 66210 / 42971 / 52435 & 40850 / 23590 / \textbf{41450} & 0.79 / 0.79 / 1.01\\
\bottomrule
\end{tabular}
\end{table*}

\noindent\textbf{Production-scale single-node ($n_h{=}128$, SP=8,
Table~\ref{tab:tpv3}).} The small-$n_h$ crossover understates \laga{}'s
production-scale position: at V3 dims the crossover \emph{flips in \laga{}'s
favor} --- \laga{} beats explicit by \textbf{1.04--1.06$\times$ at all
sequence lengths} ($S{=}4096/8192/16384$) under the fused kernel. The reason
is structural --- explicit's K/V all-to-all scales with $n_h$ (4$\times$
larger at $n_h{=}128$ vs 32), and \laga{}'s halved communication outweighs
the kernel overhead. This is the throughput-face confirmation of
\S\ref{sec:comm}: the advantage strengthens at production scale.

\begin{table*}[t]
\caption{Single-node throughput at production scale (SP=8, $n_h{=}128$, V3 dims).
Eager vs fused attention. Under fused, \laga{} wins at all seq lengths and
the absorbed memory trap is measured cleanly (no eager score-matrix confound).}
\label{tab:tpv3}
\centering\footnotesize
\setlength{\tabcolsep}{4pt}
\begin{tabular}{c|cccc|cccc}
\toprule
& \multicolumn{4}{c|}{eager attention} & \multicolumn{4}{c}{fused attention}\\
seq & explicit & absorbed & \laga{} (\laga{}/ex) & mem ex/B2/\laga{} & explicit & absorbed & \laga{} (\laga{}/ex) & mem ex/B2/\laga{}\\
\midrule
4096  & 43359 & 32489 & \textbf{45214} (1.04$\times$) & 2917/3493/2929 & 53060 & 38373 & \textbf{55571} (\textbf{1.05$\times$}) & 1796/3387/1808\\
8192  & 34770 & 22171 & \textbf{36239} (1.04$\times$) & 7842/10204/7857 & 51673 & 28839 & \textbf{54883} (\textbf{1.06$\times$}) & 2214/7541/2231\\
16384 & 20238 & OOM & \textbf{20771} (1.03$\times$) & 27348/---/27374 & 38273 & 16767 & \textbf{39932} (\textbf{1.04$\times$}) & 3044/22281/3071\\
\bottomrule
\end{tabular}
\\[2pt]
\raggedright\footnotesize Memory in MB. Fused eliminates the shared
$(B,n_h/P,S,S)$ score matrix: explicit/\laga{} peak drops from
$\sim$27\,GB to $\sim$3\,GB at $S{=}16384$, exposing absorbed's
method-specific intermediates ($+$19.2\,GB inflation vs $+$9.2\,GB under
eager, where the score matrix masked part of the gap).
\end{table*}

\noindent\textbf{Multi-node --- the comm-bound regime where \laga{} is meant
to win (Table~\ref{tab:tpmn}).} We re-run at V3 dims on \textbf{2 nodes
$\times$ 8 cards (SP=16, cross-node over RoCE $\approx 11$\,GB/s)}.

\begin{figure}[t]
\centering
\includegraphics[width=\columnwidth]{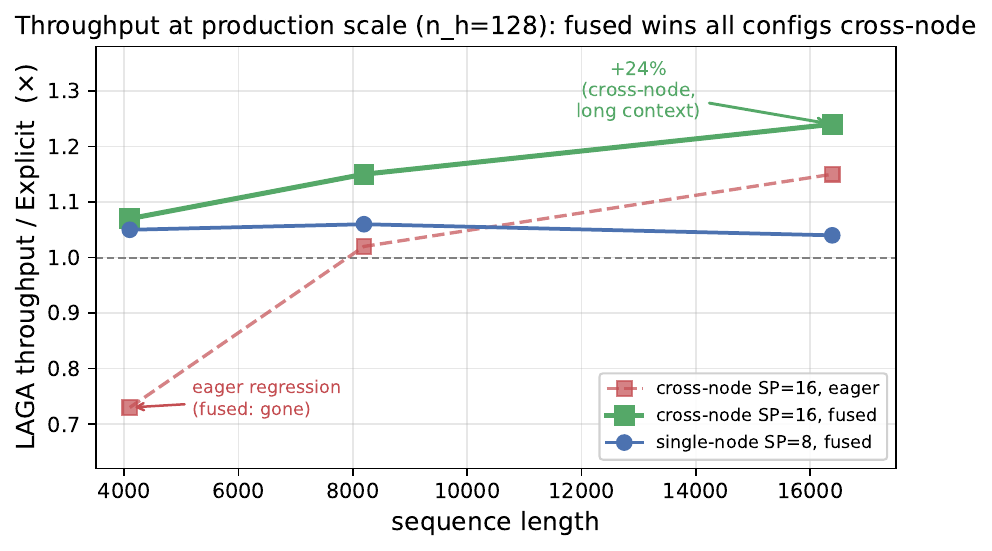}
\caption{\textbf{Key result.} \laga{}/Explicit throughput ratio cross-node
(SP=16, $n_h{=}128$). Eager: \laga{} loses short (0.73$\times$ at $S{=}4096$),
wins long ($+$15\%). Fused: the short-seq regression vanishes --- \laga{}
wins at \emph{all} seq lengths (1.07--1.24$\times$), rising to $+$24\%.}
\label{fig:tp}
\end{figure}

\begin{table*}[t]
\caption{Multi-node throughput (tokens/s), V3 dims, SP=16 cross-node.
Eager vs fused. Under fused, \laga{} wins at all sequence lengths; the eager
short-seq regression (0.73$\times$) was a kernel artifact.}
\label{tab:tpmn}
\centering\footnotesize
\begin{tabular}{c|ccc|c|ccc|c}
\toprule
& \multicolumn{4}{c|}{eager attention} & \multicolumn{4}{c}{fused attention}\\
seq & explicit & absorbed & \laga{} & \laga{}/ex & explicit & absorbed & \laga{} & \laga{}/ex\\
\midrule
4096  & 37761 & 37913 & 27538 & 0.73$\times$ & 43247 & 38980 & \textbf{46273} & \textbf{1.07$\times$}\\
8192  & 42225 & 33724 & 43254 & 1.02$\times$ & 56369 & 43048 & \textbf{65067} & \textbf{1.15$\times$}\\
16384 & 30202 & 20807 & 34728 & 1.15$\times$ & 50623 & 30895 & \textbf{62866} & \textbf{1.24$\times$}\\
\bottomrule
\end{tabular}
\end{table*}

The cross-node run isolates the inter-node effect at fixed $n_h{=}128$.
Under the \emph{fused} kernel, \laga{} leads at \textbf{all} sequence lengths
cross-node (1.07--1.24$\times$, Table~\ref{tab:tpmn}), with the advantage
\emph{growing} with sequence length: $+$7\% at $S{=}4096$,
$+$15\% at $S{=}8192$, and $+$24\% at $S{=}16384$. The eager-kernel run,
by contrast, showed a short-sequence regression (0.73$\times$ at
$S{=}4096$) --- this was an artifact of the eager implementation's
materialized score matrix and small-shard kernel-launch overhead, not a
property of \laga{}: under fused attention it vanishes entirely
(\S\ref{sec:fused}). Long-context training at $S{=}16384$ \emph{requires}
scaling out to SP=16, and \laga{} is fastest there under both kernels.
\textbf{\laga{} for long context ($\geq$8--16K) --- wins single-node and
cross-node under both kernels; under fused, wins at all seq lengths
cross-node.} Absorbed is memory-disqualified regardless
(Table~\ref{tab:mem}).

\subsection{Numerical equivalence to Explicit (B1)}
\label{sec:numeq}
Because \laga{}'s step~4 is the $H_r$ slice of B1's $W_{kv\_b}(\textrm{latent})$
--- same parameters, same matmul, reordered views --- the two are numerically
identical up to floating-point accumulation order. We verify (fp32):
\begin{itemize}\itemsep0pt
\item \textbf{SP=1: bit-identical.} Output $\max|\Delta|=0.000$;
$W_q/W_{kv\_b}/W_o$ gradients $=0.000$; $h/W_{kv\_a}/\textrm{LN}$ gradients
$\leq 2{\times}10^{-5}$ (backward accumulation noise, fp32).
\item \textbf{SP=2/4/8: equivalent to floating-point precision.} After
SP-consistent gradient aggregation: output $\max|\Delta|\leq 5.6{\times}10^{-6}$
and $W_{kv\_b}$ grad $\max|\Delta|\leq 7.7{\times}10^{-4}$ across
SP$\in\{2,4,8\}$. (\laga{} and B2, sharing the einsum kernel path, agree to
the \emph{same} tolerance as each other --- the residual is generic
kernel-rounding, not an artifact of \laga{} itself.)
\end{itemize}

\begin{figure}[t]
\centering
\includegraphics[width=\columnwidth]{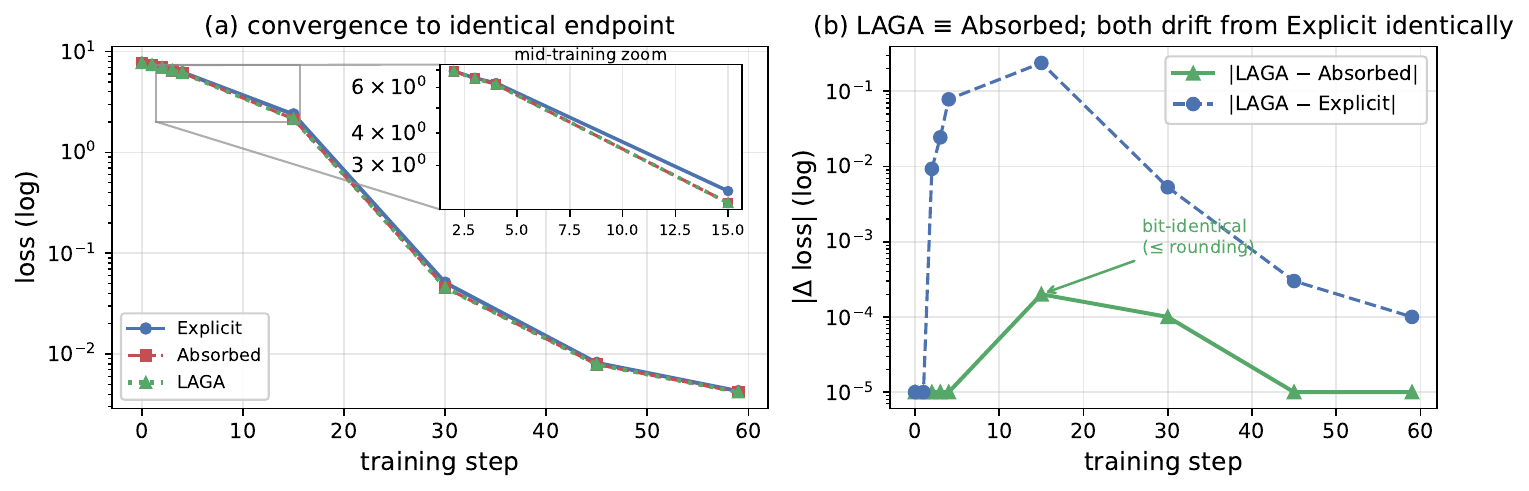}
\caption{\textbf{Correctness.} (a) Convergence to an identical endpoint.
(b) $|\Delta\textrm{loss}|$ vs Explicit: $|\laga{}-\textrm{Absorbed}|\approx 0$
(\laga{} $\equiv$ Absorbed, bit-identical) while $|\laga{}-\textrm{Explicit}|$
is the generic fp-chaos.}
\label{fig:correct}
\end{figure}

\noindent\textbf{Convergence (multi-step training).} We train a small MLA LM
(4 layers, $d{=}512$, $n_h{=}16$, seq=512, Adam lr=$10^{-3}$, 60 steps,
random-data memorization) under all three forms from an identical seed. All
three descend from loss 7.711 to $\approx 0.004$ (identical endpoint).
Transient mid-training trajectory differences (max $|\laga{}-\textrm{explicit}|
\approx 1.5$) are the well-known signature of optimizer chaos under
floating-point kernel reordering --- \laga{} and B2 are
\textbf{bit-identical to each other at every step} and diverge from explicit
by identical amounts, confirming \laga{} introduces no pathology beyond the
generic fp-equivalence class.

\subsection{Fused-kernel validation}
\label{sec:fused}
The throughput results above use an eager softmax
($\textrm{scores}{=}QK^{\top}\!\to\!\textrm{softmax}\!\to\!@V$), which
materializes the $(B,n_h/P,S,S)$ score matrix and dispatches small per-head
GEMMs. To confirm our findings are not artifacts of this implementation, we
replace it with a fused attention kernel --- \texttt{npu\_fusion\_attention}
on Ascend (which natively supports MLA's asymmetric head dimensions,
$D_q{=}D_k{=}192\!\geq\!D_v{=}128$) --- and re-run single-node (SP=8) and
cross-node (SP=16). Tables~\ref{tab:tpv3}--\ref{tab:tpmn} report both
side-by-side. Three findings:

\noindent\textbf{(F1) The eager OOM was implementation-specific, not
method-specific.} At $S{=}16384$, absorbed OOMs under eager but runs at
16.8K toks/s under fused (score matrix no longer materialized).

\noindent\textbf{(F2) The short-sequence cross-node regression was a kernel
artifact.} Under eager, \laga{} lost 0.73$\times$ cross-node at $S{=}4096$;
under fused, it \emph{wins} 1.07$\times$. The regression was the eager
score-matrix overhead and per-shard kernel-launch cost --- not a property of
\laga{}. Under fused, \laga{} leads at all sequence lengths cross-node
(1.07--1.24$\times$).

\noindent\textbf{(F3) The memory trap is measured more cleanly under fused.}
Eliminating the shared $(B,n_h/P,S,S)$ score matrix removes a confound:
explicit/\laga{} peak memory drops from $\sim$27\,GB to $\sim$3\,GB at
$S{=}16384$, so absorbed's method-specific intermediates are isolated. Its
inflation grows from $+$9.2\,GB (eager) to $+$19.2\,GB (fused) --- the
absorbed $q_{\textrm{absorbed}}$/accumulator still cost $n_h d_{kv}$ per
token regardless of kernel, but under eager the score matrix masked part of
the gap. Under fused the C1 measurement is unconfounded.

\noindent\textbf{Takeaway.} The communication advantage (C2/C3) and the
memory-trap measurement (C1) are both robust to kernel choice. The fused
results strengthen the paper: \laga{} wins at all sequence lengths
cross-node, and the memory trap is exposed without the eager confound.

\subsection{Limitations and scope}
\label{sec:limits}
\begin{itemize}\itemsep0pt
\item \textbf{Prototype scale.} Comm/memory are measured on a single
attention layer and throughput on a 4-layer stack at V3 head dimensions, not
an end-to-end DeepSeek-V3-scale (61-layer) run. Comm is per-layer and
analytical, and the memory-trap mechanism ($n_h d_{kv}$ intermediates) is
structural and independent of depth; however, end-to-end wall-clock and MFU
at full model depth (where attention's share of total compute is smaller)
remain future work.
\item \textbf{Convergence at toy scale.} The multi-step convergence check
(\S\ref{sec:numeq}) uses a 4-layer, $d{=}512$ model on a random-data
memorization task (60 steps). We rely on the per-forward bit-identical
result (SP=1: output and $W_q/W_{kv\_b}/W_o$ gradients all exactly $0$) as
the primary correctness guarantee; the convergence run only confirms no
optimizer pathology, not realistic-training parity.
\item \textbf{No direct MCore comparison.} Throughput is measured against our
B1 (which reproduces MCore's explicit MLA math), not MCore itself; a
head-to-head against MCore on matched hardware is left to future work (MCore
runs on CUDA, our prototype on NPU).
\item \textbf{Fused up-projection kernel.} Our fused attention replaces the
score-matrix computation but not \laga{}'s local per-head up-projection
einsums; a single fused up-proj$\to$attention kernel would further amortize
the small-shard GEMMs. Future work.
\item \textbf{Training only.} \laga{} targets training (no KV cache);
inference-time MLA parallelism is the domain of TPLA/MLRA/Helix.
\item \textbf{Two-node.} Multi-node uses two nodes; larger cross-node CP is
the natural next scale-up.
\end{itemize}

\subsection{Cross-hardware validation (NVIDIA A100)}
\label{sec:xhw}
To confirm the structural results are not a hardware artifact, we re-run the
V3-scale single-node benchmark (SP=8, $n_h{=}128$) on NVIDIA A100
(CUDA/NCCL) via the same code path.
\begin{itemize}\itemsep0pt
\item \textbf{Communication (byte-identical).} Comm is analytical, so it
matches exactly across hardware: \textbf{1.98$\times$} at all three sequence
lengths.
\item \textbf{Memory (the trap reproduces).} Absorbed inflation reproduces
on A100 ($+$575/$+$2221/$+$8734\,MB at $S$=4K/8K/16K) vs Ascend
($+$575/$+$2362/$+$9247\,MB) --- agreement within allocator noise ($\leq$6\%).
\item \textbf{Throughput (crossover shape holds).} \laga{}/ex $=$
1.05/1.03/1.02$\times$ on A100 vs 1.04/1.04/1.03$\times$ on Ascend --- the
crossover shape (mild \laga{} lead at all $S$) reproduces across hardware.
\textbf{Absorbed OOMs at $S{=}16384$ on both cards} (Ascend eager 36.6\,GB
peak; A100 44.7\,GB peak), while explicit/\laga{} run --- direct evidence
that the memory trap, not the eager-kernel limit, disqualifies absorbed.
\end{itemize}

\begin{table*}[t]
\caption{Cross-hardware agreement (V3, SP=8, $n_h{=}128$). Comm is
byte-identical; the absorbed memory trap and the \laga{} throughput win
reproduce on CUDA.}
\label{tab:xhw}
\centering\footnotesize
\begin{tabular}{l|c|cc}
\toprule
metric & $S$ & Ascend 910B & NVIDIA A100\\
\midrule
comm reduction (ex/\laga{})    & 4K/8K/16K & 1.98$\times$/1.98$\times$/1.98$\times$ & 1.98$\times$/1.98$\times$/1.98$\times$\\
absorbed mem inflation          & 4K/8K/16K & $+$575/$+$2362/$+$9247\,MB & $+$575/$+$2221/$+$8734\,MB\\
\laga{}/ex throughput          & 4K/8K/16K & 1.04$\times$/1.04$\times$/1.03$\times$ & 1.05$\times$/1.03$\times$/1.02$\times$\\
\bottomrule
\end{tabular}
\end{table*}

\section{Correctness}
\label{sec:correctness}
\laga{} computes the same function as Explicit (B1); they differ only in
\emph{where the up-projection happens} (before vs after the collective) and
\emph{which tensor crosses} (per-head K/V vs latent).

\subsection{Equivalence to Explicit}
\label{sec:equiv}
Let $L = \textrm{LN}(W_{kv\_a}h)$. Explicit computes, for every head $i$ and
token $t$,
\[
k_{\textrm{nope}}[i,t]=W_{kv\_b}^{k}[i]\,L[t],\qquad
v[i,t]=W_{kv\_b}^{v}[i]\,L[t],
\]
via a single batched matmul. \laga{}'s step~4 computes exactly the same,
for $i\in H_r$, via per-shard einsum. These are the \emph{same linear maps
applied to the same tensor}, differing only in kernel dispatch order. Hence:

\noindent\textbf{Proposition 1 (exact equivalence, SP=1).} \emph{With $P=1$,
\laga{} and Explicit produce identical outputs and parameter gradients up
to floating-point reordering of the $W_{kv\_b}$ vs per-shard einsum kernels.}

\noindent\textbf{Proposition 2 (floating-point equivalence, SP$>$1).}
\emph{For $P>1$, the sole difference is that Explicit all-to-alls per-head K/V
while \laga{} all-gathers the latent; both are reorderings of the same
arithmetic, so outputs and aggregated gradients agree to floating-point
precision.}

\subsection{Backward and the $W_{kv\_b}$ gradient shard}
\label{sec:backward}
The collectives are wrapped as autograd functions: the head-scatter
all-to-all's backward is the inverse (seq-scatter) all-to-all; the sequence
all-gather's backward is a reduce-scatter. The $W_{kv\_b}$ gradient
partitions differ: B1 holds a (local\_seq, all\_heads) shard; \laga{} holds
a (full\_seq, $H_r$) shard; both sum to the true total via one SP-group
all-reduce.

\subsection{Decoupled RoPE}
The $k_{pe}$ band is shared across heads (P2) and never compressed, so it
bypasses the absorb question: \laga{} all-gathers $k_{pe}$ alongside the
latent. The rope term $q_{\textrm{rope}}\!\cdot\!k_{pe}$ is identical across
all three forms.

\section{Composition with Expert Parallelism}
\label{sec:ep}
MoE models (e.g., DeepSeek-V3) interleave MLA attention with MoE FFNs trained
under expert parallelism (EP). \textbf{Collective isolation:} \laga{}'s
collectives operate on head/seq axes inside the attention block; EP's
dispatch/combine all-to-alls operate on the token/expert axis inside the MoE
block --- temporally separated, never contending. \textbf{Sequence-then-token
partitioning:} SP first partitions the sequence; EP then dispatches those
tokens via the standard preprocess. \laga{} places no constraint on the EP
degree. \textbf{Gradient aggregation:} the $W_{kv\_b}$ head-axis all-reduce is
independent of EP's data-parallel reduction; the two compose by summation.

\section{Related Work}
\label{sec:related}
\noindent\textbf{Megatron-Core MLA (most direct comparison).} MCore
~\cite{megatroncore} ships a single MLA training class whose forward is the
explicit form (our B1); the absorb reformulation is gated to inference and
hard-asserted against training. \textbf{Our \S\ref{sec:trap}/\ref{sec:mem}
result explains why that restriction is well-founded} --- the absorbed form
is a training-memory trap (C1) --- and \laga{} closes the gap it leaves
(C2): a low-communication MLA \emph{training} path. We are not aware of any
shipped path --- in MCore or elsewhere --- that does latent-over-CP
$+$ local up-projection for training.

\noindent\textbf{Distributed MLA (inference).}
TPLA~\cite{tang2026tpla}, MLRA~\cite{liu2026mlra}, and
Helix~\cite{helix2025} parallelize MLA for \emph{decoding/serving}. All
exploit the inference KV-cache; none addresses training activation memory.
\laga{} targets training.

\noindent\textbf{Sequence parallelism (MHA/GQA).}
Megatron-SP~\cite{korthikanti2023megatronsp}, Ulysses~\cite{jacobs2023ulysses},
Ring/Striped Attention~\cite{liu2024ringattention,striped2023},
USP~\cite{usp2024}, LoongTrain~\cite{loongtrain2024} assume per-head KV and
do not evaluate MLA. Under P1 their head-scatter has nothing to scatter.

\noindent\textbf{Name disambiguation.} ``LASP'' (linear-attention SP,
Sun et al.~\cite{sun2024lasp}; LASP-2~\cite{sun2025lasp2}) is
orthogonal: it targets \emph{linear attention}, not softmax MLA. Our method
is \laga{} (Latent All-Gather Attention) to avoid that collision.

\noindent\textbf{Expert parallelism.} MegaBlocks~\cite{gale2023megablocks},
Tutel~\cite{hwang2023tutel}, Lina~\cite{li2023lina}, DeepEP~\cite{deepep}
provide EP dispatch all-to-alls; \laga{}'s collectives compose with EP
(\S\ref{sec:ep}).

\noindent\textbf{MLA training-memory analyses.} ``Memory Analysis on Training
DeepSeek Models''~\cite{memanalysis2025} analyzes MLA activation memory under
TP/SP/CP in Megatron-LM; it does not compare the explicit/absorbed/\laga{}
patterns on the memory axis we isolate, nor report the absorbed inflation.

\section{Discussion and Limitations}
\label{sec:discuss}
\noindent\textbf{When to choose \laga{} (the crossover).} \laga{} wins when
the workload is communication-bound: long sequences, moderate SP, slow
interconnects, cross-node SP groups --- the long-context training scenario
MLA is deployed for. Explicit wins when compute-bound.

\noindent\textbf{Eager implementation caps sequence at low SP.} Our prototype
uses eager softmax; a fused FA-style kernel over the scores lifts the cap
(future work; orthogonal to the comm/memory comparisons).

\noindent\textbf{Fused up-projection for the compute-bound regime.} A single
fused up-proj$\to$attention$\to$output kernel would amortize the local
up-projection matmuls and is the route to closing the crossover gap.

\noindent\textbf{Training scope.} \laga{} targets training, where there is no
KV cache. Inference-time MLA parallelism is out of scope.

\noindent\textbf{A generalizable lesson.} The absorbed-baseline result is
broader than MLA: an optimization designed for the inference KV-cache becomes
a \emph{memory regression} in training. Practitioners porting inference
optimizations to training should check activation footprint, not just
communication.

\noindent\textbf{MoE routing amplifies matmul nondeterminism (a
reproducibility note).} An incidental observation: the top-$k$ routing gate
is a \emph{discrete amplifier} of floating-point noise --- a matmul with
nondeterministic reduction order produces a router score near a tie; top-$k$
flips which expert a token visits; the output jumps discretely, turning
sub-ULP noise into a self-sustaining divergence. In a controlled two-run,
same-seed prototype (identical init/data, no dropout), MoE run-to-run loss
divergence exceeded a same-architecture dense control by $\approx 16\times$,
appearing as a single $\approx 1000\times$ step at exactly the first
routing-flip step, and deterministic mode eliminated it. The effect is
\emph{hardware-contingent}: it requires the backend to supply matmul
nondeterminism --- present on the NPU (aclnn default reduction order), latent
on CUDA where cuBLAS matmul and the relevant scatter/index ops are
deterministic at our prototype scale (the effect did not reproduce there). We
report this as a reproducibility caveat for NPU MoE training --- same-seed
runs diverge cliff-shaped rather than gradually, and checkpoint-resume
fidelity should be judged by loss range/trend, not bit-equality --- not as a
large-scale result, and leave full characterization to future work.

\noindent\textbf{Open directions.} (i) A fused \laga{} attention kernel;
(ii) extending \laga{} to ring-style SP for the $P>n_h$ regime;
(iii) overlapping the latent all-gather with the preceding sub-layer's compute.

\bibliography{references}
\bibliographystyle{mlsys2026}

\end{document}